\documentclass[letterpaper,aps,prd,10.5pt,preprintnumbers,showpacs,showkeys,superscriptaddress,nofootinbib,amsmath,amssymb,floatfix
]{revtex4}
\usepackage{times}
\usepackage{amsmath}
\usepackage{amssymb}
\usepackage{amsfonts}
\usepackage{mathrsfs}
\usepackage{graphicx}
\usepackage{subfigure}
\usepackage{booktabs}
\usepackage{rotating}
\usepackage{longtable} 
\usepackage{bm}
\usepackage{float}
\usepackage{epstopdf}
\usepackage{comment}
\usepackage{graphicx}
\usepackage{subfigure}
\usepackage{amsmath}

\makeatletter

\newcommand{\Rmnum}[1]{\expandafter\@slowromancap\romannumeral #1@}
\makeatother

\makeatletter
\newcommand*{\rom}[1]{\expandafter\@slowromancap\romannumeral #1@}
\makeatother

\makeatletter

    \usepackage[colorlinks]{hyperref} 
\hypersetup{
  colorlinks   = true, 	
  urlcolor     = cyan, 	
  linkcolor    = blue, 	
  citecolor   = magenta 
} 
\usepackage[dvipsnames]{xcolor}

\begin{document}
\title{Accretion Disk Luminosity and Topological Characteristics for a Schwarzschild Black Hole
Surrounded by King Dark Matter Halo}
\author{Soroush Zare}
\email{szare@uva.es}
\affiliation{Departamento de F\'{\i}sica Te\'orica, At\'omica y Optica and Laboratory for Disruptive \\ Interdisciplinary Science (LaDIS), Universidad de Valladolid, 47011 Valladolid, Spain}
\author{Farokhnaz Hosseinifar}
\email{f.hoseinifar94@gmail.com}
\affiliation{Center for Theoretical Physics, Khazar University, 41 Mehseti Street, Baku, AZ-1096, Azerbaijan}
\author{L.M. Nieto}
\email{luismiguel.nieto.calzada@uva.es}
\affiliation{Departamento de F\'{\i}sica Te\'orica, At\'omica y Optica and Laboratory for Disruptive \\ Interdisciplinary Science (LaDIS), Universidad de Valladolid, 47011 Valladolid, Spain}
\author{Dhruba Jyoti Gogoi}
\email{moloydhruba@yahoo.in}
\affiliation{Department of Physics, Moran College, Moranhat, Charaideo 785670, Assam, India.}
\affiliation{Theoretical Physics Division, Centre for Atmospheric Studies, Dibrugarh University, Dibrugarh
786004, Assam, India.}
\author{Kuantay Boshkayev}
\email{kuantay@mail.ru}
\affiliation{Al-Farabi Kazakh National University, Al-Farabi ave. 71, 050040 Almaty, Kazakhstan}
\author{Ainur Urazalina}
\email{y.a.a.707@mail.ru}
\affiliation{Al-Farabi Kazakh National University, Al-Farabi ave. 71, 050040 Almaty, Kazakhstan}
\author{Hassan  Hassanabadi}
\email{hassan.hassanabadi@uva.es}
\affiliation{Departamento de F\'{\i}sica Te\'orica, At\'omica y Optica and Laboratory for Disruptive \\ Interdisciplinary Science (LaDIS), Universidad de Valladolid, 47011 Valladolid, Spain}
\affiliation{Department   of   Physics, Faculty of Science,   University   of   Hradec   Kr\'{a}lov\'{e},  Rokitansk\'{e}ho 62, 500   03   Hradec   Kr\'{a}lov\'{e},   Czechia}


\begin{abstract}

\hspace{10cm}
\\
\\
\begin{normalsize}
\textbf{Abstract}\\
\\

This study delves into the intricate properties of a Schwarzschild black hole enveloped by King dark matter in an isotropic configuration.
The thermodynamic characteristics of this black hole are meticulously analyzed, and the dynamics of massive and massless particles in its vicinity are investigated. In examining the trajectories of massless particles, the shadow cast in the presence of King dark matter is explored, revealing virtual ranges for the corresponding parameters. For the dynamics of massive particles, the radius of the innermost stable circular orbit, angular momentum, energy, and angular velocity of a test particle within the King dark matter framework surrounding the black hole are calculated.
The effect of King dark matter on the accretion disk energy flux, effective radiation temperature, differential luminosity, and spectral luminosity are then investigated. The stability of the photon sphere in the presence of King dark matter is also studied, and finally, the thermodynamic potentials of this black hole are examined from a topological perspective.

\end{normalsize}

\bigskip\bigskip

\end{abstract}

\maketitle

\section{Introduction}\label{Sec1}

Black holes have long been one of the most intriguing phenomena in astrophysics \cite{novikov1973astrophysics,novikov2013physics,heckman2014coevolution}. Initially, they were thought to be regions of space devoid of matter, where not even light could escape their gravitational pull \cite{ruffini1971introducing,penrose1972black,rees1974black,novikov1995black}. However, subsequent observations have revealed that black holes are surrounded by a rich environment composed of various forms of matter \cite{chapline1975cosmological,schneider2002first,cyburt2003primordial,volonteri2012black,carr2016primordial}, including gas \cite{lynden1969galactic,shakura1973black}, dust \cite{antonucci1993unified,urry1995unified}, and, most intriguingly, dark matter \cite{zwicky1933rotverschiebung,rubin1980rotational,blumenthal1984formation,trimble1987existence}.
This revelation has transformed our understanding of the dynamics of black holes and the cosmos \cite{hawley2005foundations,volonteri2012black}: dark matter, which does not emit or interact with electromagnetic radiation like ordinary matter, is hypothesized to constitute a substantial part of the total mass of the universe \cite{turner1991dark,overduin2004dark,spergel2015dark,bertone2018history}.

As researchers strive to unravel the complexities of dark matter, numerous different models have been proposed to explain its properties, distribution, and interactions, especially in the vicinity of black holes. \cite{wang2016dark,oks2021brief,shen2024analytical}. Many scientists have contributed to this field by suggesting various models that describe the behavior of dark matter and its influence on the gravitational dynamics around black holes. Some of these notable models, each with its own characteristics and implications for our understanding of the universe, are the Navarro-Frenk-White \cite{navarro1996structure} profile, the Hernquist \cite{hernquist1990analytical} profile, the Burkert \cite{burkert1995structure} profile, the Moore \cite{moore1994evidence} profile, and the Dehnen \cite{dehnen1993family} profile, among the most relevant.
The effects of dark matter surrounding black holes have been investigated from a multitude of perspectives, encompassing thermodynamic properties \cite{xu2019perfect,singh2021thermodynamic,pantig2023black,carvalho2023thermodynamics,gohain2024thermodynamics}, quasinormal modes \cite{cardoso2016black,jusufi2020quasinormal,bamber2021quasinormal,konoplya2021black,das2023stability,konoplya2025quasinormal}, optical appearance \cite{konoplya2019shadow,jusufi2019black,saurabh2021imprints,figueiredo2023black,capozziello2023dark,chowdhury2025effect}, and beyond \cite{frampton2010primordial,ivanov1994inflation,schulze2011effect,xu2018black,kavanagh2020detecting,konoplya2019hawking}.

One of the most exciting areas of black hole research is the study of accretion disks, which refers to the rotating disk that accretes around a black hole \cite{thorne1974disk,pringle1981accretion,abramowicz2013foundations,li2002accretion,yuan2014hot,nieto2025accretion}. The intense gravitational pull of the black hole causes surrounding material to fall toward it. \cite{narayan2005black,yuan2014hot,porth2017black}. Key quantities studied in the context of accretion disks include flux and luminosity, which provide insight into the energy output and efficiency of the accretion process \cite{abramowicz1989thick,balbus1998instability,calvet1999evolution,harko2009can,klessen2010accretion}.
The influence of dark matter on accretion disks around black holes has received significant attention \cite{hatziminaoglou2001accretion,howell2008dark,pugliese2022dark,boshkayev2022accretion,kurmanov2022accretion}.
For instance, Ref. \cite{boshkayev2020accretion} explores how dark matter affects the luminosity, energy flux, and other related quantities of these systems, offering a comparative analysis between rotating and non-rotating black holes embedded in dark matter halos. Furthermore, Ref. \cite{hui2023black} explores the dynamics of black hole accretion in the presence of dark matter, focusing on how dark matter influences the mass and angular momentum of black holes through a process called `over-superradiance'. This phenomenon allows the mass of the accretion cloud to exceed typical limits.

Investigation of the dynamics of a Schwarzschild black hole surrounded by King dark matter has multiple motivations. From a theoretical perspective, it helps us to generalize the classical results of black hole physics, such as horizon properties, thermodynamic properties, optical appearances, Hawking radiation, and geodesic structure, by considering the influence of the King dark matter distribution. On the observation side, the effects of the King dark matter halo can appear in the measurable effects by the different observatories, such as shifts in the shadow of a black hole, photon sphere radius, and some evidence about accretion efficiency.

This article, in which we delve into the dynamics of a Schwarzschild black hole enveloped by King dark matter \cite{king1962structure,kar2025diverse}, is organized as follows. 
Section \ref{Sec2} is dedicated to the calculation of the horizon radius, the Hawking temperature and the subsequent determination of the radius of the black hole remnant.
In Section \ref{Sec3}, we explore the trajectories of massless particles in the black hole vicinity, identifying the photon sphere and the black hole shadow. From observational data, we derive constraints on the relevant parameters. 
In Section \ref{Sec4}, we turn our attention to the behaviour of massive particles, analysing energy flux, spectral luminosity, and efficiency.
In Sections \ref{Sec6} and \ref{Sec7}, we investigate the stability and instability of the photon sphere, along with a classification employing black hole thermodynamic potentials, approached through a topological perspective. Finally, Section \ref{Sec8} summarizes our findings.

\section{Thermodynamic Properties of a Schwarzschild black hole in the presence of King dark matter}\label{Sec2}
\begin{figure}[htb]
	\centering
  \includegraphics[width=6.4cm]{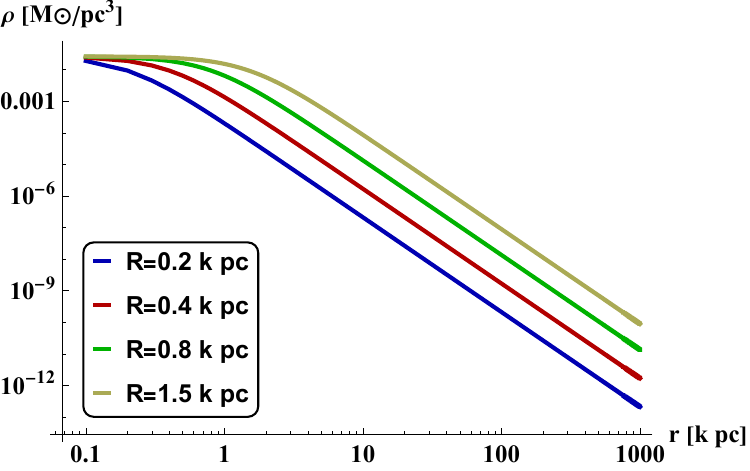} 
    \caption{Density profile \eqref{den} for $\rho_0=1/(37.96 ) M_\odot/pc^3$ and several values of $R$.}\label{fig:Den}
\end{figure}

\begin{figure}[htb]
	\centering
  \includegraphics[width=6.4cm]{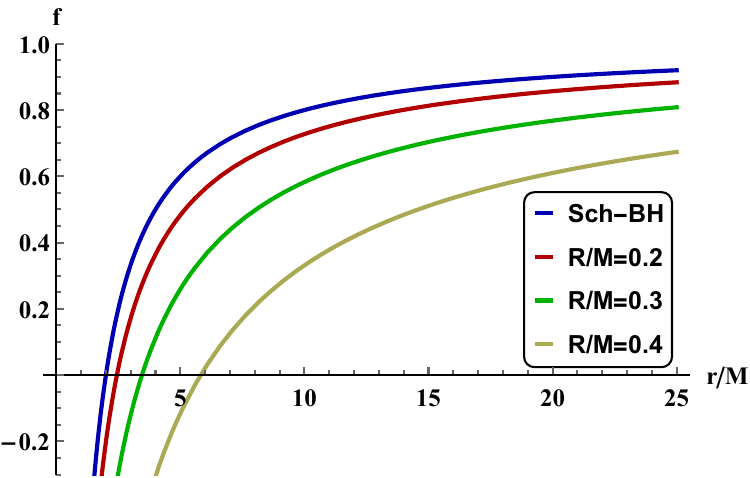} \hspace{-0.2cm}\\
    \caption{Variation of the lapse function in terms of $r/M$ for $\rho_0 M^2=1$.}
    \label{fig:fr}
\end{figure}

\begin{figure}[htb]
	\centering
\includegraphics[width=6.4cm]{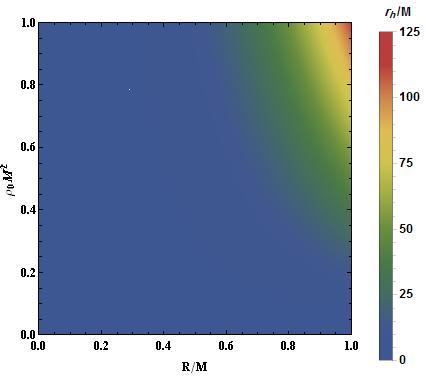}
    \caption{Variation of horizon radius $r_h$ in terms of $R/M$ and $\rho_0 M^2$. Increasing either $R/M$ or $\rho_0 M^2$ enhances the magnitude of the horizon radius.}\label{fig:horizon}
\end{figure}

In the present study, in which we adopt the metric signature $(-+++)$ and set $\hbar=G=c=1$, we investigate certain properties of a Schwarzschild black hole in the presence of King dark matter \cite{king1962structure}, whose metric is elegantly derived in reference \cite{kar2025diverse}.
This model is suitable for describing the distribution of matter in galactic halos and can effectively characterize the gravitational behavior and structure of galaxies in various regions. Furthermore, the King model has been selected as a preferred option for analyses related to black holes and galactic halos due to its mathematical simplicity and compatibility with astronomical observations \cite{konoplya2022solutions}.
In \cite{kar2025diverse}, a spherically symmetric spacetime described by
\begin{eqnarray}
ds^2= -f(r) dt^2+\frac{dr^2}{g(r)}+h(r) (d\theta^2+\sin^2\theta d\phi^2),\label{ds2}
\end{eqnarray}
where
\begin{equation}\label{hr}
f(r)=g(r)=1-2\frac{m(r)}{r},\qquad  h(r)=r^2,
\end{equation}
and $m(r)$ represents the mass function, is considered, and Einstein's equations are solved, formulating the Einstein field equations as 
\begin{equation}\label{Ein}
R_{\mu\nu} - \frac{1}{2} R g_{\mu\nu} = 8\pi T_{\mu\nu}.
\end{equation}
Here, $g_{\mu\nu}$ is the spacetime metric, $R_{\mu\nu}$ is the Ricci tensor, $R$ is the Ricci scalar, and $T_{\mu\nu}$ is the energy-momentum tensor defined as $T_{\mu\nu}=\text{diag}[-\rho,\,\tilde{\tau},\,p,\,p]$, where $\rho$ represents the energy density, $\tilde{\tau}$ denotes the radial pressure, and $p$ is the tangential pressure. 
Through a series of meticulous calculations, the following results are obtained
\begin{equation}\label{para}
\rho = -\tilde{\tau}=\frac{\partial_r m(r)}{4\pi r^2}\quad\text{and} \quad p=-\frac{\partial_r^2 m(r)}{8\pi r}.
\end{equation}
The parameterized Dekel-Zhao density profile \cite{zhao1996analytical}, given by 
\begin{equation}\label{denDZ}
\rho(r)=\rho_0\frac{\left(\dfrac{r}{R}\right)^{\mu-3}}{\left(1+\left(\dfrac{r}{R}\right)^{\nu}\right)^{\frac{\mu+\alpha}{\nu}}},
\end{equation}
where $R$ represents the scale radius, $\rho_0$ is the central density, and $(\mu,\, \nu,\, \alpha)$ are dimensionless parameters of the density profile, is then used.
Specifically, the King dark matter density profile is obtained for the values $(3,\,2,\,0)$ in the Dekel-Zhao profile \cite{king1962structure}, such that we have
\begin{equation}\label{den}
\rho(r)=\rho_0\left(1+\left(\frac{r}{R}\right)^2\right)^{-3/2}.
\end{equation}
This density has also been investigated under the designation of Beta model \cite{sofue2020rotation,kurmanov2023analysis}.
For clarity, Figure~\ref{fig:Den} shows the 
variation of the density profile \eqref{den} as a function 
of $r$ for different values of $R$ and a certain initial value of 
$\rho_0$.
\\
Next, using Eq.~\eqref{hr}, the time delay function is calculated in \cite{kar2025diverse} and turns out to be
\begin{equation}
f(r)=1 -\frac{2 M}{r}+\frac{8 \pi  \rho_0 R^3}{\sqrt{r^2+R^2}}+\frac{8 \pi  \rho_0 R^3}{r} \ln \left(\frac{\sqrt{r^2+R^2}-r}{R}\right).\label{fr}
\end{equation}
From this expression it is evident that in the limits of $\rho_0\to 0$ or $R\to 0$, the time-lapse function $f(r)$ approaches the Schwarzschild black hole solution. The behavior of the curve $f(r)$ for $\rho_0 M^2=1$ is shown in Figure~\ref{fig:fr}, 
where it can be clearly seen that $f(r)$ has a root that determines the radius of the black hole horizon.
Furthermore, as can be seen in Fig. \ref{fig:fr} and also from Eq.~\eqref{fr}, an increase in $R/M\,(\rho_0 M^2)$ results in an enlargement of the black hole horizon radius which is obtained from $f(r_h)=0$. Fig. \ref{fig:horizon} illustrates the variations in the horizon radius in terms of changes in the parameters $R/M$ and $\rho_0 M^2$.

The mass of the black hole is expressed as a function of the horizon radius $r_h$ from $f(r_h)\big|_{M=M_h}=0$ as
\begin{eqnarray}\label{Mh}
M_h(r_h)=\frac{r_h}{2} +\frac{4 \pi  \rho_0 r_h R^3}{\sqrt{r_h^2+R^2}}+4 \pi  \rho_0 R^3 \ln \left(\frac{\sqrt{r_h^2+R^2}-r_h}{R}\right),
\end{eqnarray}
which in the limits of $R\to 0$ or $\rho_0\to 0$ corresponds to the mass of the Schwarzschild black hole.

\begin{figure}[ht!]
	\centering
  \includegraphics[width=6.4cm]{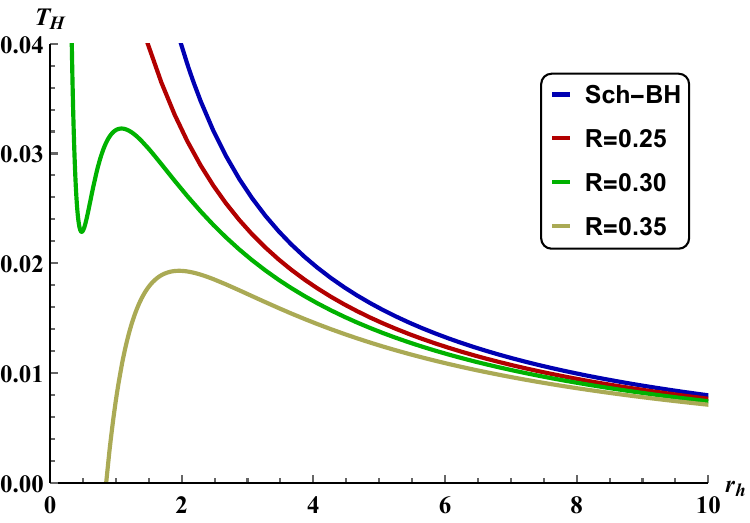} \hspace{-0.2cm}\\
    \caption{Hawking temperature versus $r_h$ for the case $\rho_0=1$. In the Hawking temperature curve of a Schwarzschild  black hole, no phase transition is observed. Increasing the effect of the parameter $R$ beyond a certain threshold leads to the existence of two phase transitions in the Hawking temperature. Increasing the size of this parameter causes the Hawking temperature of the black hole to become zero at some point, indicating a remnant radius, and in this case, only one phase transition is observed in the Hawking temperature of the black hole.}\label{fig:TH}
\end{figure}

The Hawking temperature is calculated from $1/(4\pi)\partial_r f(r)\big|_{r=r_h}$ and using Eq.~\eqref{Mh} is expressed as \cite{hawking1974black}
\begin{eqnarray}\label{TH}
T_{\rm H}=\frac{1}{4 \pi  r_h}-\frac{2 \rho_0 r_h R^3}{\left(r_h^2+R^2\right)^{3/2}},
\end{eqnarray}
indicating that at $R\to 0$ or $\rho_0\to 0$, it approaches the Hawking temperature of the Schwarzschild black hole. 
The Hawking temperature is plotted as a function of the horizon radius in Figure~\ref{fig:TH}.
It is clear that for certain parameter choices, one or two second-order phase transition can be observed in the black hole temperature curve at $\partial_{r_h} T_{\rm H}\big|_{r_h=r_c}=0$. We are particularly interested in states that feature a single extreme (a maximum),  leading  to the appearance of a remnant radius in the black hole.

\begin{figure}[ht!]
	\centering
  \includegraphics[width=6.4cm]{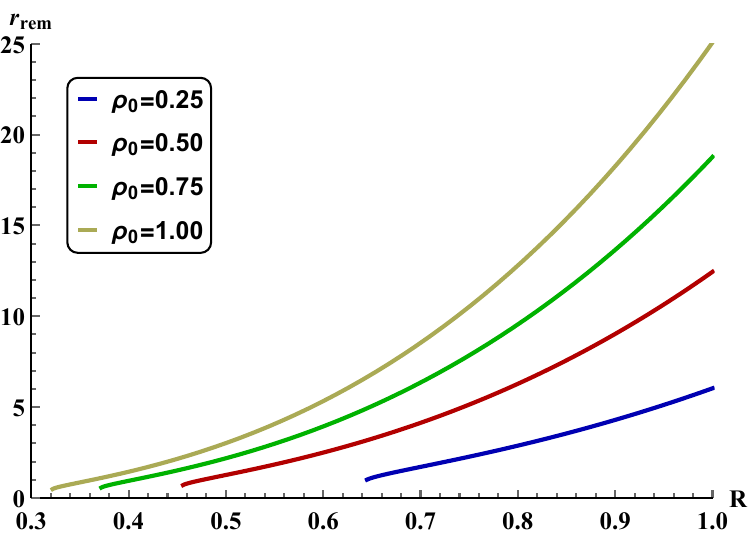} \hspace{0.9cm}
  \includegraphics[width=7.1cm]{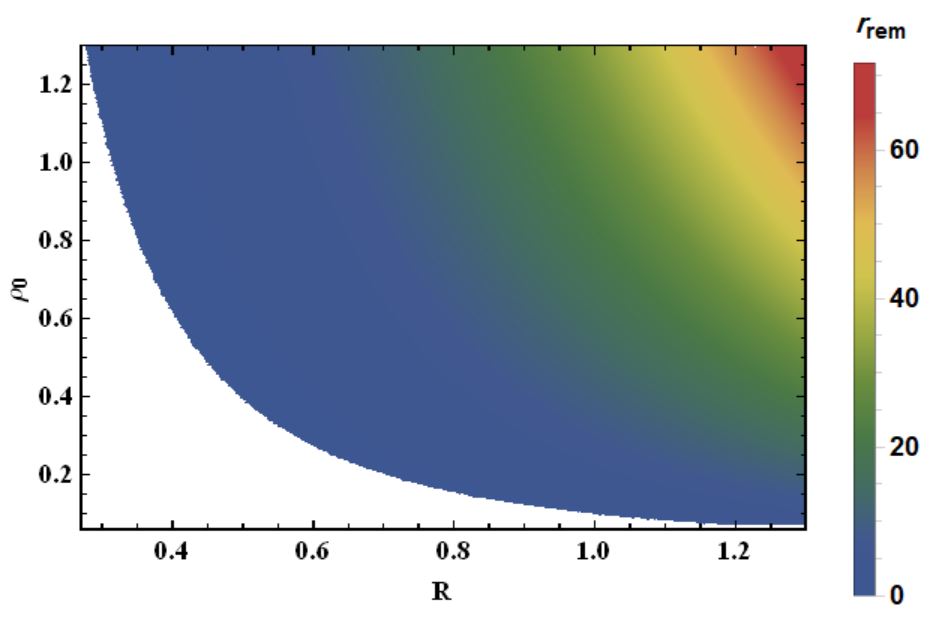} \hspace{-0.2cm}\\
    \caption{Left panel: Remnant radius curves for varying $R$. Right panel: The colored area indicates the range of variables for which a remnant radius exist.}\label{fig:rRem}
\end{figure}

The remnant radius $r_{\text{rem}}$ refers to the radius  remaining after the evaporation of the black hole \cite{giddings1992black,koch2005black} and is calculated from $T_{\rm H}\big|_{r_h=r_{\text{rem}}}=0$ \cite{chen2024quasi}. As illustrated in Fig. \ref{fig:TH}, the existence of the remnant radius depends on  the choice of parameters. Fig. \ref{fig:rRem} describes the variations in the remnant radius as a function of the parameter $R$, as well as the range of parameters $R$ and $\rho_0$ for which a remnant radius (and consequently a second-order phase transition in the Hawking temperature) is observed.
As can be seen from this figure, increasing $R\,(\rho_0)$ produces an increase in the magnitude of the remaining radius.

\section{Trajectory of massless particles around a Schwarzschild black hole surrounded by King dark matter}\label{Sec3}

The study of black hole shadows and the paths of light rays near them has gained great importance, particularly following the Event Horizon Telescope (EHT) observations \cite{akiyama2019first,chael2021observing,johnson2023key}, because it provides important insights into the properties and behavior of black holes. By comparing the black hole shadow   with EHT observations, a range of black hole parameters can be identified for which the shadow is within an acceptable range \cite{kumar2020black,afrin2023eht,raza2024shadow}. 
To study the shadow, we first examine the black hole's photon sphere, where light can orbit the black hole in circular paths. As light rays approach this region, they can either escape to infinity or fall into the black hole. This boundary creates a disk in the observer's sky, which is interpreted as the black hole's shadow  \cite{perlick2022calculating,solanki2022photon}.
To obtain the shadow in the spherically symmetric spacetime introduced in the Eq. \eqref{ds2}, for $\theta=\pi/2$, the Lagrangian $\mathcal{L}(x,\,\dot{x})=\frac{1}{2}g_{\mu\nu}\dot{x}^\mu\dot{x}^\nu$ is written as \cite{chandrasekhar1998mathematical,gibbons2015jacobi,shaikh2019shadows}
\begin{eqnarray}\label{LagSh}
\mathcal{L}(x,\,\dot{x})=\frac{1}{2}\left(-f(r)\dot{t}^2+\frac{\dot{r}^2}{f(r)}+h(r)\dot{\phi}^2\right),
\end{eqnarray}
where $x^\mu=(t,\,r,\,\phi)$, $\dot{x}^{\mu}$ are the derivatives of the coordinates, and $g_{\mu\nu}$ represents the metric tensor of the spacetime. The Euler-Lagrange equations are expressed as
\begin{eqnarray}\label{dLagsh}
\frac{d}{d\lambda}\left(\frac{\partial\mathcal{L}}{\partial\dot{x}^\mu}\right)-\frac{\partial\mathcal{L}}{\partial x^\mu}=0,
\end{eqnarray}
where $\lambda$ is an affine parameter along the geodesic. 
Using the equations related to the $t$ and $\phi$ components, two constant of motion can be derived as \cite{perlick2018black}
\begin{eqnarray}\label{MotionCons}
E=\frac{\partial\mathcal{L}}{\partial\dot{t}},\qquad L =\frac{\partial\mathcal{L}}{\partial\dot{\phi}}.
\end{eqnarray}
Thus, the equation of motion for the radial component is given by \cite{li2024geodesic}
\begin{align}\label{trajj}
\frac{1}{2}\dot{r}^2+\frac{1}{2}\left(\delta+\frac{L^2}{h(r)}\right)f(r)=\frac{1}{2}E^2,
\end{align}
where for light-like particles $\delta=0$ and for time-like particles $\delta=1$.

\begin{figure}[ht!]
	\centering
  \includegraphics[width=6.4cm]{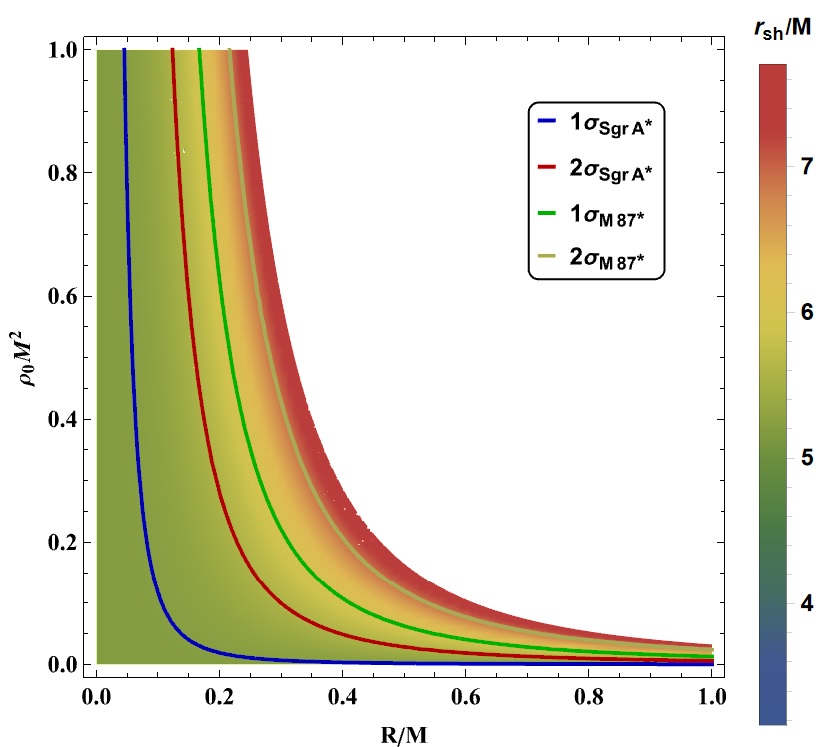} \hspace{-0.2cm}
    \caption{Density of shadow radius by varying $R/M$ and $\rho_0 M^2$. The colored lines represent the upper limit of the permissible regions $1\sigma$ and $2\sigma$ based on EHT observations for Sgr A$^*$ and M 87$^*$.}\label{fig:shadow}
\end{figure}

For light-like particles, considering $V^l_{\rm eff}(r)=f(r)L^2/h(r)$, Eq. \eqref{trajj} can be expressed as
\begin{align}\label{trajsh}
\dot{r}^2+V^l_{\rm eff}(r)=E^2.
\end{align}
Due to the circular orbit, the following restriction must be satisfied in the photon sphere \cite{virbhadra2000schwarzschild}
\begin{eqnarray}\label{photon}
\dot{r}=0\Longrightarrow L^2\frac{f(r_{\rm ph})}{h(r_{\rm ph})}=E^2\qquad\text{and}\qquad
\ddot{r}=0\Longrightarrow \partial_{r_{\rm ph}}\frac{f(r_{\rm ph})}{h(r_{\rm ph})}=0.
\end{eqnarray}
Using the above equation, the black hole photon radius, $\color{red}r_{\rm ph}$, would be found. Furthermore, assuming $L/E$ is fixed, for the asymptotically flat time-lapse function of the form Eq. \eqref{fr}, the black hole shadow radius $\color{red}r_{\rm sh}$ is given by \cite{bozza2010gravitational}
\begin{equation}\label{shadow}
\begin{aligned}
r_{\rm sh}&=\frac{r_{\rm ph}}{\sqrt{f(r_{\rm ph})}}\\
&=\frac{\sqrt{r_{\rm ph}^3 \sqrt{R^2+r_{\rm ph}^2}}}{\sqrt{-2 M \sqrt{R^2+r_{\rm ph}^2}+8 \pi  \rho_0 R^3 r_{\rm ph}+r_{\rm ph} \sqrt{R^2+r_{\rm ph}^2}+8 \pi  \rho_0 R^3 \sqrt{R^2+r_{\rm ph}^2} \left[\log \left(\sqrt{R^2+r_{\rm ph}^2}-r_{\rm ph}\right)-\log (R)\right]}}.
\end{aligned}
\end{equation}
Due to EHT observations for Sgr A$^*$, it is known that the acceptable range for the shadow radius in the $1\sigma$ and $2\sigma$ regions are $4.55<r_{\rm sh}/M<5.22$ and $4.21<r_{\rm sh}/M<5.56$, respectively \cite{vagnozzi2023horizon}. Similarly, for M 87$^*$ the shadow radius for the $1\sigma$ and $2\sigma$ regions should be within the ranges $4.26<r_{\rm sh}/M<6.03$ and $3.38<r_{\rm sh}/M<6.91$, respectively \cite{akiyama2019first}.
Therefore, by comparing the black hole shadow calculated from Eq. \eqref{shadow} with the boundaries of the $1\sigma$ and $2\sigma$ regions, we can set upper and lower bounds on the parameters that would allow the shadow to fall within these ranges.
Figure \ref{fig:shadow} illustrates the variation of the shadow radius when varying the parameters $R/M$ and $\rho_0 M^2$ in the interval $(0,\,1)$.
In the limit where $R$ or $\rho_0$ approach zero, the shadow of the black hole in the presence of King dark matter corresponds to the shadow of the Schwarzschild black hole  $3\sqrt{3}M$. 
As the parameter $R$ or $\rho_0$ increases, the magnitude of the black hole's shadow also expands. Therefore, there is no lower bound on the parameters that keeps the shadow within the permissible region, although we can calculate upper bounds.
We observe that by selecting the parameter $\rho_0 M^2$ within the range of 0.1 to 0.2, the upper limit for the parameter $R/M$ for the regions $1\sigma$ and $2\sigma$ in the case of Sgr A$^*$ is $0.08<R/M<0.11$ and $0.23<R/M<0.30$, respectively, while for M 87$^*$, these values are $0.31<R/M<0.41$ and $0.41<R/M<0.54$, respectively.

\section{Radiative Flux and Spectral 
Luminosity of the Accretion Disk for a Schwarzschild black hole surrounded by King dark matter}\label{Sec4}

In the previous section, the path of light for massless particles was examined. In this section, to extract the luminosity of accretion disks around black holes, we will investigate the trajectory of massive particles. To study the dynamics of massive particles in the gravitational field, we employ a model similar to the one in the previous section. According to Eq. \eqref{trajj}, the effective potential for timelike particles is given by \cite{fathi2022study}
\begin{eqnarray}\label{traj}
V^t_{\text{eff}}(r)=\frac{f(r)}{2}\left(1+\frac{L^2}{h(r)}\right),
\end{eqnarray}
and Eq. \eqref{trajj} can be expressed as
\begin{eqnarray}
\frac{1}{2}\dot{r}^2+V^t_{\rm eff}(r)=\frac{1}{2}E^2.
\end{eqnarray}
The study of the circular orbits of particles in the accretion disk is crucial for calculating the luminosity.
For a marginally stable circular trajectory denoting the ISCO radius, which is the closest stable circular orbit that massive particles can maintain around a black hole, the following conditions must be satisfied \cite{shaymatov2021effect,shen2025inner,jusufi2025black}
\begin{eqnarray}\label{rdot}
\dot{r}=0\longrightarrow V^t_{\text{eff}}(r)=\frac{1}{2}E^2,\quad \ddot{r}=0\longrightarrow \partial_r V^t_{\text{eff}}(r)=0,\quad\text{and}\quad \dddot{r}=0\longrightarrow\partial^2_r V^t_{\text{eff}}(r)=0.
\end{eqnarray}
Using Eq. \eqref{rdot} the ISCO radius is calculated from
\begin{equation}\label{ISCO}
\bigg(h'(r) \left(2 f(r) f'(r) h'(r)+h(r) \left(f(r) f''(r)-2 f'(r)^2\right)\right)-f(r) h(r) f'(r) h''(r)\bigg)_{r=r_{\rm ISCO}}=0.
\end{equation}
In the context of our study, if ISCO is present, all circular orbits beyond ISCO are stable.
Table \ref{Table:Radii} represents the ISCO radius of the black hole in the presence of King dark matter, along with the corresponding values for the horizon radius, the photon radius, and the shadow.
\begin{table}[ht!]
	\caption{Impact of King dark matter on black hole radii.}
	\centering
	\label{Table:Radii}
	\begin{tabular}{|c|c|c|c|c|c|}
		\hline
		$\rho_0 M^2$& $R/M$ & $r_h/M$ & $r_{\rm ph}/M$ & $r_{\rm sh}/M$ & $r_{\rm ISCO}/M$\\
		\hline
		$0.0$ & $0.0$ & $2.00000$ & $3.00000$ & $5.19615$ & $6.00000$\\
		\hline
		$0.5$ & $0.1$ & $2.03402$ & $3.05240$ & $5.29780$ & $6.09330$\\
		\hline
		 $0.5$ & $0.2$ & $2.21135$ & $3.32816$ & $5.85313$ & $6.56837$\\
		\hline
		 $0.5$ & $0.3$ & $2.63659$ & $3.99546$ & $7.23151$ & $7.71933$\\
		\hline
		$1.0$ & $0.1$ & $2.06847$ & $3.10544$ & $5.40064$ & $6.18806$\\
		\hline
		 $1.0$ & $0.2$ & $2.44246$ & $3.68694$ & $6.56673$ & $7.20502$\\
		\hline
		 $1.0$ & $0.3$ & $3.45365$ & $5.26965$ & $9.77510$ & $10.03397$\\
		\hline
	\end{tabular}
\end{table}
It is clear that the presence of King dark matter, along with an increase in either parameter $R$ or $\rho_0$, results in an enlargement of all these radii. The presence of dark matter affects on the gravitational field for test particles, strong attraction of the halo increases the magnitude of ISCO radius.

\begin{figure}[ht!]
\centering
  \includegraphics[width=6.4cm]{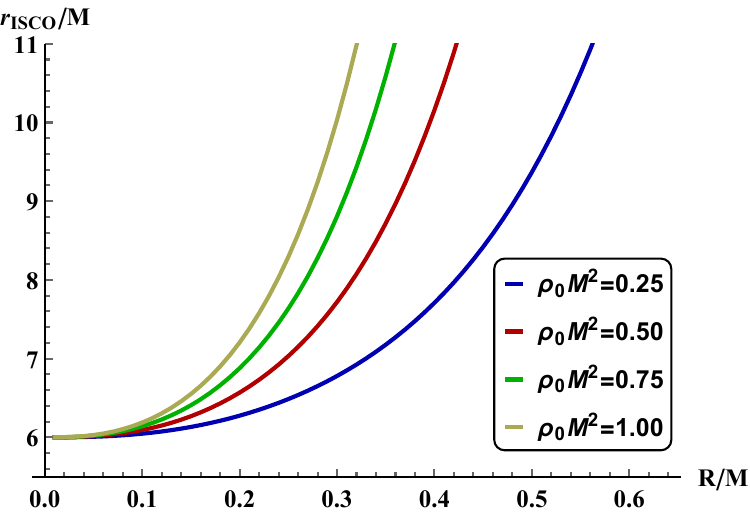} \hspace{0.9cm}
  \includegraphics[width=6.4cm]{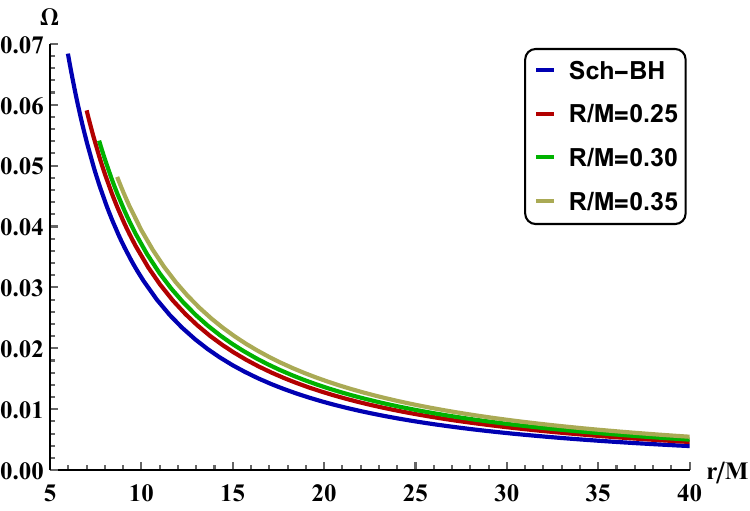} \hspace{-0.2cm}\\[2ex]
  \includegraphics[width=6.4cm]{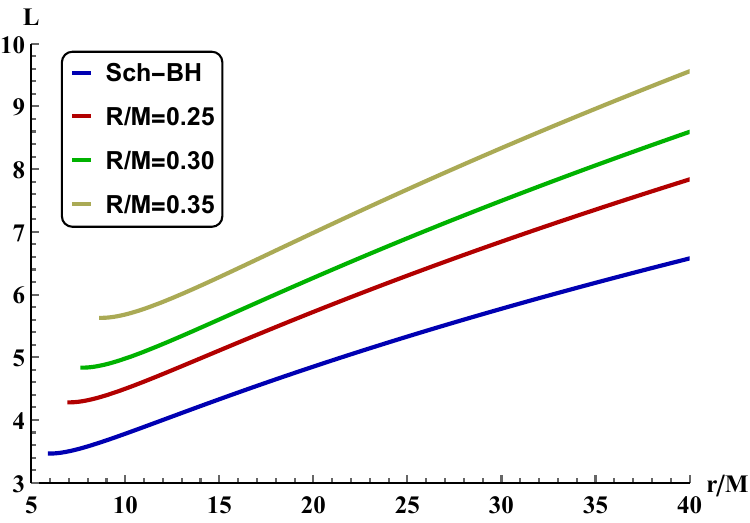} \hspace{0.9cm}
  \includegraphics[width=6.4cm]{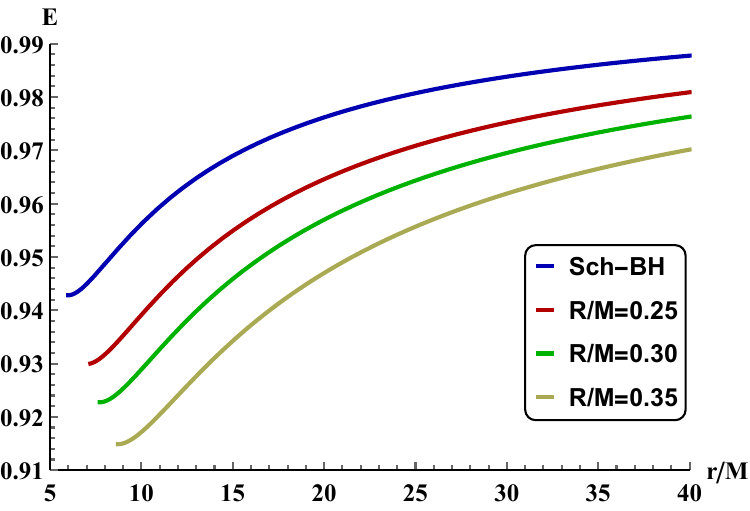} \hspace{-0.2cm}\\
    \caption{Variation of $r_{\text{ISCO}}$ in terms of $R/M$, and curves of angular velocity, angular momentum and energy as a function of $r/M$ considering $\rho_0 M^2=0.5$. All this for the Schwarzschild black hole and three cases of $R/M$.}\label{fig:AC1}
\end{figure}

On the other hand, the specific angular velocity of the test particles in the accretion disk using the equation \eqref{rdot} is derived from \cite{boshkayev2021luminosity}\begin{align}\label{omega}
\Omega(r)&=\sqrt{\frac{\partial_r f(r)}{\partial_r h(r)}}=\frac{1}{r}\sqrt{\frac{M+4 \pi  \rho_0 R^3 \left(\ln \left(\sqrt{r^2+R^2}+r\right)-\ln R\right)}{r}-\frac{4 \pi  \rho_0 R^3 \left(2 r^2+R^2\right)}{\left(r^2+R^2\right)^{3/2}}}.
\end{align}
Furthermore, using the equations \eqref{rdot} and \eqref{omega}, the specific energy of a test particle and the specific angular momentum are calculated from
\begin{equation}
\label{Er}
\begin{aligned}
E(r)&=\frac{f(r)}{\sqrt{f(r)+\Omega^2(r) h(r)}}
\\&=
\dfrac{-2 M \sqrt{r^2+R^2}+r \left(\sqrt{r^2+R^2}+8 \pi  \rho_0 R^3\right)+8 \pi  \rho_0 R^3 \sqrt{r^2+R^2} \left(\ln \left(\sqrt{r^2+R^2}-r\right)-\ln (R)\right)}{r \sqrt{\left(r^2+R^2\right) \left(1-\dfrac{3 M}{r}+\dfrac{4 \pi  \rho_0 R^3 \left(4 r^2+3 R^2\right)}{\left(r^2+R^2\right)^{3/2}}-\dfrac{4 \pi  \rho_0 R^3 \left(-2 \ln \left(\sqrt{r^2+R^2}-r\right)+\ln \left(\sqrt{r^2+R^2}+r\right)+\ln (R)\right)}{r}\right)}}
,
\end{aligned}
\end{equation}
and
\begin{equation}
\label{Lr}
\begin{aligned}
L(r)&=\frac{\Omega(r)h(r)}{\sqrt{f(r)+\Omega^2(r) h(r)}}
\\&=
\frac{r \sqrt{\dfrac{M}{r}-\dfrac{4 \pi  \rho_0 R^3 \left(2 r^2+R^2\right)}{\left(r^2+R^2\right)^{3/2}}+\dfrac{4 \pi  \rho_0 R^3 \left(\ln \left(\sqrt{r^2+R^2}+r\right)-\ln (R)\right)}{r}}}{\sqrt{1-\dfrac{3 M}{r}+\dfrac{4 \pi  \rho_0 R^3 \left(4 r^2+3 R^2\right)}{\left(r^2+R^2\right)^{3/2}}-\dfrac{4 \pi  \rho_0 R^3 \left(-2 \ln \left(\sqrt{r^2+R^2}-r\right)+\ln \left(\sqrt{r^2+R^2}+r\right)+\ln (R)\right)}{r}}}
.
\end{aligned}
\end{equation}
Figure \ref{fig:AC1} shows the ISCO radius  ($r_{\text{ISCO}}$) curves and the behavior of the angular velocity ($\Omega$), specific angular momentum  ($L$) and specific energy ($E$).
In the limit of $R\to 0$ or $\rho_o\to 0$, the black hole in Eq. \eqref{ds2} resembles the Schwarzschild black hole, with its ISCO radius $r_{\rm ISCO sch-BH}=6M$. Increasing the parameters $R/M$ and $\rho_0 M^2$ results in an expansion of the ISCO. Due to the changes of gravitational, at larger radii, increasing $R/M$ or $\rho_0 M^2$ by a fixed amount has a more significant influence on the ISCO size than at smaller radii. 
Furthermore, the impact of increasing the parameter $R/M$ on the quantities $\Omega$, $E$, and $L$ is illustrated compared to the Schwarzschild black hole. It is observed that, in the presence of King dark matter, the test particles in the accretion disk exhibit higher angular velocity and angular momentum, while possessing lower energy.  Specifically, it can be stated that at small values of $r/M$, the presence of King dark matter leads to an increase in mass at a fixed radius, resulting in a stronger gravitational field experienced by test particles. Consequently, the value of angular velocity becomes larger in the presence of King dark matter. However, as $r/M$ increases, $\Omega(r)$ approaches zero due to the reduction in gravity at greater distances. Also, Angular momentum is greater with King dark matter than without at the same $r/M$, because dark matter increases gravitational attraction, which requires particles to have more angular momentum to balance this force and avoid falling into the black hole. Thus, in the presence of King dark matter, particles must move faster to maintain their orbits. On the other hand, in relativistic mechanics, $E(r)$ represents the energy per unit mass of a particle. At infinity, specific energy reaches a maximum of $1$, while for bound orbits it becomes less. A smaller $E(r)$ indicates a more strongly bound particle and in the presence of King dark matter, particles are more tightly bound. It is also evident that, as the parameter $R/M$ increases, the discrepancy with the Schwarzschild black hole becomes more pronounced. For example, the energy of the test particles in the accretion disk reaches a high value for the Schwarzschild black hole, while it reaches a lower value for the case $R=0.35/M$ at a constant radius $r/M$.

\begin{figure}[ht!]
	\centering
  \includegraphics[width=6.4cm]{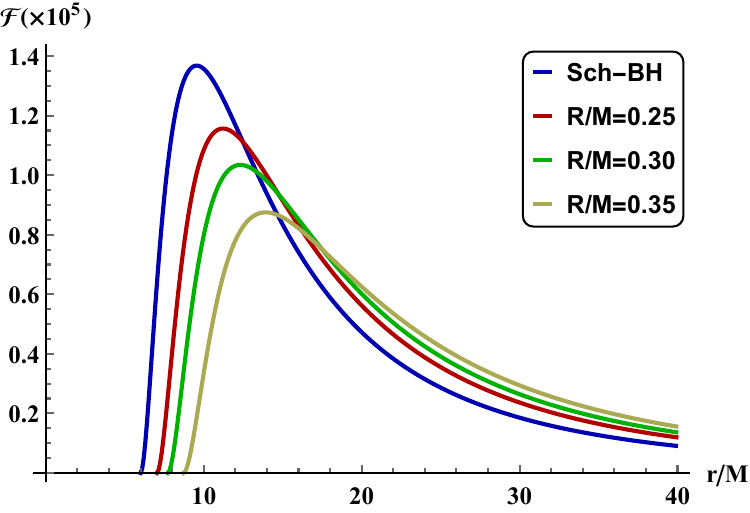} \hspace{0.9cm}
  \includegraphics[width=6.4cm]{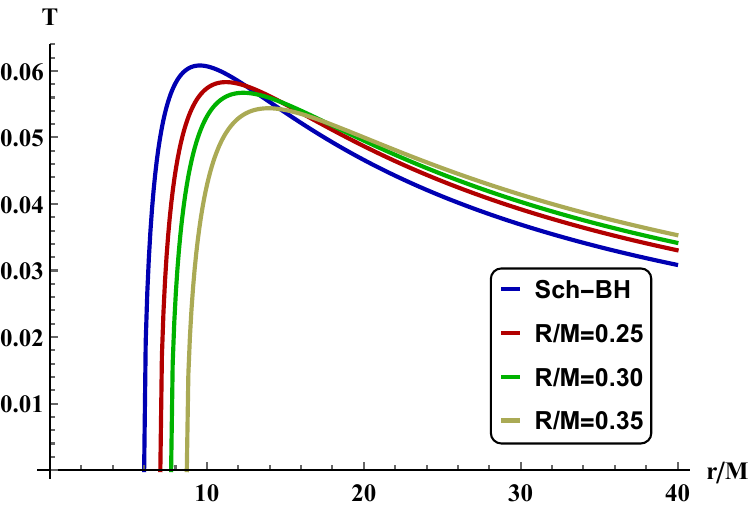} \hspace{-0.2cm}
    \caption{Left panel: Energy flux with variation of $r/M$. Right panel: Behavior of the radiation temperature in terms of $r/M$. In both cases it is assumed that $\rho_0 M^2=0.5$. }\label{fig:AC2}
\end{figure}

Using Eqs. \eqref{ISCO}--\eqref{Lr} the radiative flux emitted by the accretion disk is computed from \cite{uktamov2024particle,boshkayev2024luminosity}
\begin{eqnarray}\label{flux}
\mathfrak{F}(r)=-\frac{\dot{m}}{4\pi\sqrt{h(r)}}\frac{\partial_r \Omega(r)}{(E(r)-\Omega(r) L(r))^2}\int_{r_{\text{ISCO}}}^{r}(E(\tilde{r})-\Omega(\tilde{r}) L(\tilde{r}))\partial_{\tilde{r}}L(\tilde{r}) d \tilde{r},
\end{eqnarray}
where $\dot{m}$ represents the mass accretion rate.
Using the Stefan-Boltzmann law, a relationship can be established between the energy flux and the radiation temperature of the disk, expressed as \cite{heydari2023thin}
\begin{eqnarray}
\mathfrak{F}(r)=\sigma_{\rm SB}T^4(r).
\end{eqnarray}
In Fig. \ref{fig:AC2}, the variations of energy flux and the radiation temperature of the disk as a function of $r/M$ are shown for $\rho_0M^2=0.5$ considering $\dot{m}=1$.
It is evident that the energy flux increases sharply at first and reaches its maximum value, before gradually decreasing. Moreover, the presence of the King Dark Matter causes the black hole described by the Eq. \eqref{ds2} to experience a smaller maximum flux compared to the Schwarzschild black hole, and this maximum occurs at a larger radius. 
Increasing the parameter $R/M$ to a constant $\rho_0 M^2$ results in a further decrease in the maximum flux. It is also observed that, for a small $r/M$, the flux in the absence of the King Dark Matter is larger, and for a constant $r/M$, increasing the parameter $R/M$ reduces the flux. However, as $r/M$ increases, the flux for states with a larger $R/M$ eventually exceeds that for states with a lower $R/M$ at a constant $r/M$. This behavior also applies to the radiation temperature.
The accretion disk surrounding a black hole resembles a glowing stove, exhibiting increased heat and brightness in areas closer to the black hole. However, the influence of King dark matter modifies this dynamic. By extending the gravity outward, it causes the innermost regions of the disk to be somewhat cooler, while the outer regions emit more radiation than what is observed for a Schwarzschild black hole.
Within the ISCO, matter spirals into the black hole along unstable paths, lacking sufficient time to convert energy into radiation. Consequently, the peak of radiation is found just outside the ISCO, where viscous processes are still functioning effectively. The presence of King dark matter alters the gravitational potential, which in turn shifts the optimal region for these processes. This results in the peak temperature and energy flux moving outward, with the maximum intensity being slightly diminished compared to the Schwarzschild case.

\begin{figure}[ht!]
	\centering
  \includegraphics[width=6.4cm]{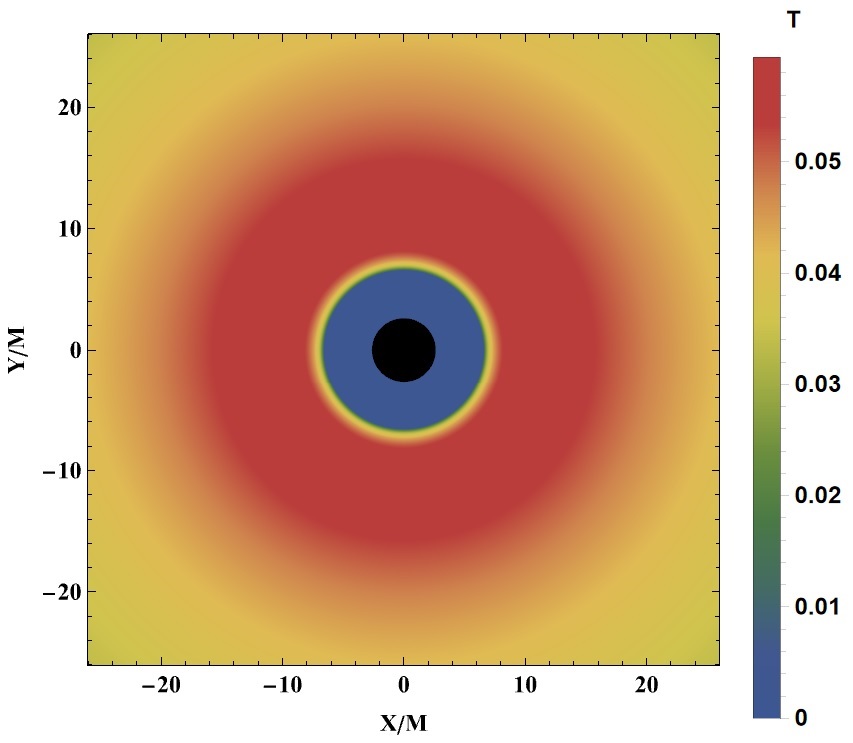} \hspace{0.9cm}
  \includegraphics[width=6.4cm]{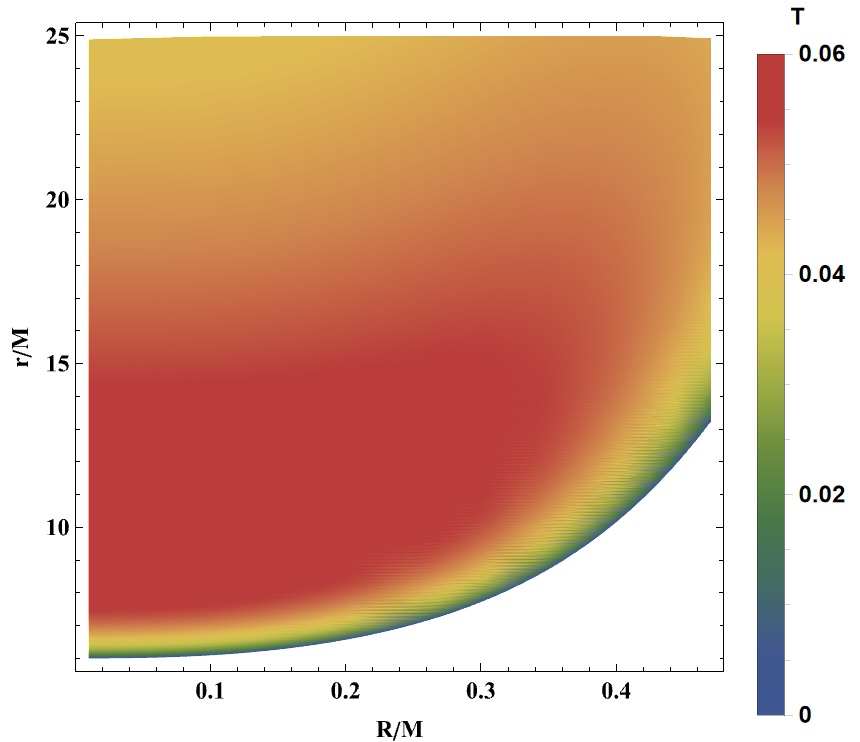} \hspace{-0.2cm}
    \caption{Radiation temperature for $\rho_0 M^2=0.5$. Left panel: Assuming $R/M=0.20$ in the equatorial Cartesian coordinate plane $X-Y$. Right panel: In terms of $R/M$ and $r/M$.}\label{fig:AC3}
\end{figure}

In Fig. \ref{fig:AC3}, the emitted temperature density is shown for the case $\rho_0 M^2=0.5$.
In the left panel, for $R/M=0.2$, the variations in radiation temperature in the $X-Y$ plane beyond the horizon are illustrated. As expected, the temperature is zero up to the ISCO radius, after which it reaches its maximum and then decreases. In the right panel, the variations in radiative temperature as a function of $r/M$ and $R/M$ are presented. 
As expected, increasing $R/M$ leads to a larger ISCO radius, which initiates temperature variations at a larger $r/M$. Additionally,  increasing $R/M$ results in a lower maximum radiation temperature, which occurs at a larger radius, consistent with the results in Fig. \ref{fig:AC2}.

\begin{figure}[ht!]
	\centering
  \includegraphics[width=6.4cm]{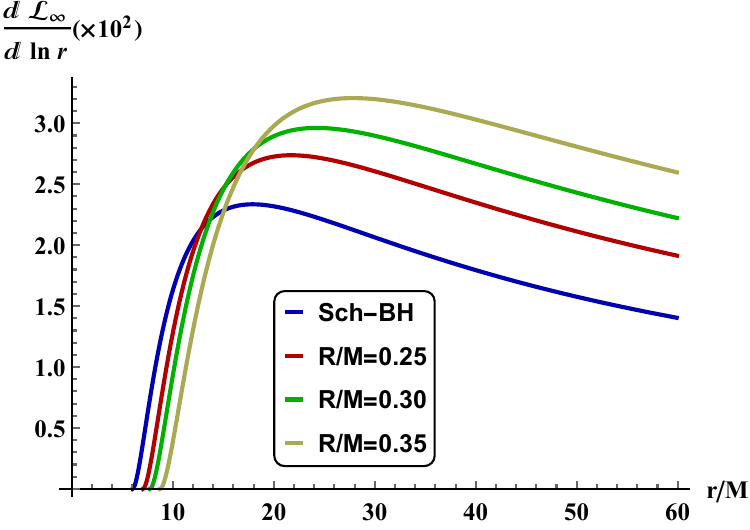} \hspace{0.9cm}
  \includegraphics[width=7cm]{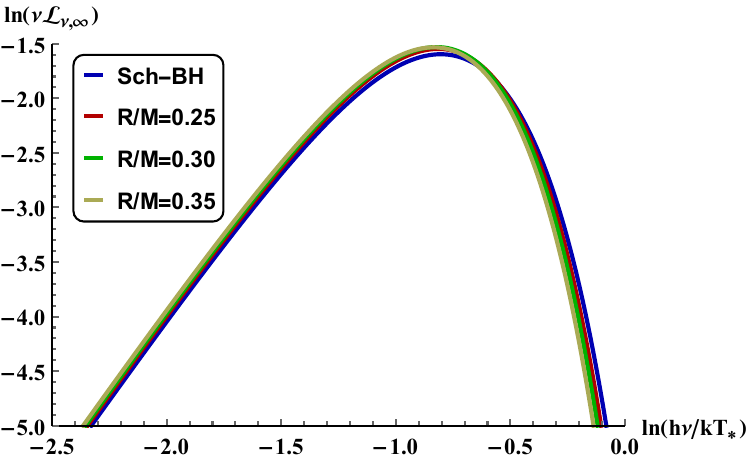} \hspace{-0.2cm}
    \caption{Left panel: Differential luminosity of the accretion disk at infinity in terms of $r/M$. Right panel: Spectral luminosity of the accretion disk as a function of the blackbody radiation frequency. The value $\rho_0 M^2=0.5$ has been chosen. }\label{fig:AC4}
\end{figure}

The differential luminosity represents the energy per unit time reaching an observer at infinity. Using the energy flux defined in Eq. \eqref{flux}, it is calculated from \cite{shaymatov2023epicyclic,alloqulov2024radiation}
\begin{eqnarray}
\frac{d \mathfrak{L}_\infty}{d \ln r}=4\pi r^2 E(r) \mathfrak{F}(r),
\end{eqnarray}
and the observed spectral luminosity distribution of the accretion disk at spatial infinity is computed from \cite{boshkayev2022accretion,alloqulov2024electric}
\begin{eqnarray}
\nu\mathfrak{L}_{\nu,\infty}=\frac{15}{\pi^4}\int_{r_i}^\infty \left(\frac{d \mathfrak{L}_\infty}{d \ln r}\right)\frac{\left(u(r)y\right)^4}{M^2_T \mathfrak{F}(r)}\frac{1}{\exp\left(\dfrac{u(r)y}{\left[M^2_T \mathfrak{F}(r)\right]^{1/4}}\right)-1}d\ln r
,
\end{eqnarray}
where $u(r)=1/\sqrt{f(r)+\Omega^2(r)h(r)}$, $y=\hbar\nu/k T_*$, $\hbar$ refers to Plank's constant, $\nu$ indicates the frequency of the emitted radiation, $k$ indicates the Boltzmann constant, $T_*=\dot{m}/\left(4\pi M^2_T\sigma_{\rm SB}\right)$, and $M_T=M+M_H$, where $M_H$ refers to the mass profile for the dark matter distribution, which using the equation \eqref{den} turns out to be
\begin{equation}
M_H=\int 4\pi r^2 \rho(r)=4 \pi  \rho_0 R^3 \left[\ln \left(\frac{\sqrt{r^2+R^2}-r}{R}\right)-\frac{r}{\sqrt{r^2+R^2}}\right].
\end{equation}
Figure \ref{fig:AC4} shows the variation of the differential luminosity and the spectral luminosity for $\rho_0 M^2=0.5$.
It is observed that the behavior of the differential luminosity is similar to that of the energy flux curve. Specifically, the magnitude of the differential luminosity at low radii for a black hole in the presence of King dark matter is smaller than that of a Schwarzschild black hole. However, at large radii, the magnitude of the differential luminosity for a black hole in the presence of King dark matter is larger than that of a Schwarzschild black hole, and dark matter changes the emission spectrum of black hole accretion disks. At small radii, disks influenced by dark matter emit less light, while at larger radii, they radiate more strongly, resulting in a broader and redder spectrum compared to Schwarzschild disks.
This alteration helps identify black holes in dense dark matter environments. The differential luminosity behaves like energy flux, showing that an increase in $R/M$ leads to a lower maximum value of $d\mathcal{L}_{\infty}/(d \ln r)$ at smaller $r/M$.
Furthermore, it is observed that the magnitude of the spectral luminosity at low frequencies is larger for the black hole in the presence of King dark matter than for the Schwarzschild black hole, while at high frequencies, the Schwarzschild black hole exhibits a larger spectral luminosity compared to the black hole in the presence of King dark matter.
With a King dark matter halo, the inner disk cools, and the outer disk brightens, shifting luminosity toward lower frequencies while suppressing higher energies. Consequently, dark matter broadens the radiation area and lowers the disk's maximum temperature, resulting in a spectrum that is redder and less intense at high frequencies. At small $r/M$, the spectral luminosity is higher with dark matter, but at larger $r/M$, Schwarzschild black holes show greater luminosity.

\begin{figure}[ht!]
	\centering
	\includegraphics[width=6.4cm]{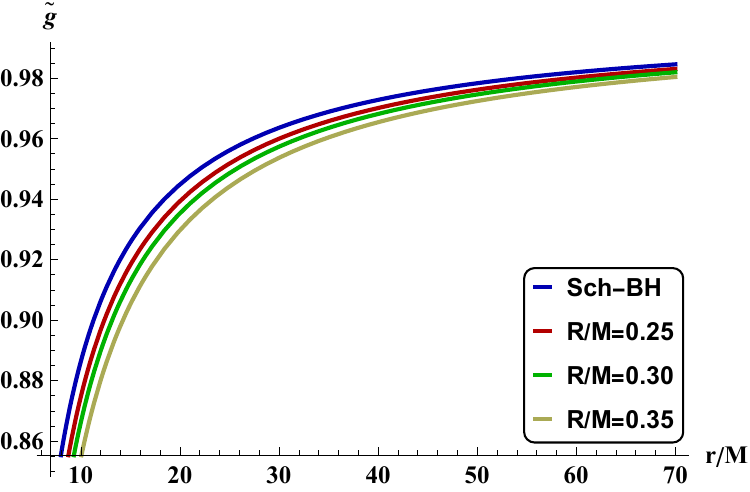} \hspace{-0.2cm}
	\caption{Redshift--factor curves in terms of $r/M$ considering $\rho_0 M^2=1$ and $\theta=\pi/8$ for the Schwarzschild black hole and in presence of King dark matter.}\label{fig:ACg}
\end{figure}

The redshift--factor which indicates the ratio of the photon that is observed by a remote observer to the emitted frequency is obtained from \cite{nozari2025investigating}
\begin{eqnarray}
	\tilde{g}(r)=\frac{1}{1+z}=\frac{\sqrt{f(r)-(r\sin\theta\Omega(r))^2}}{
		1+\Omega(r)r\sin\theta\sin\psi}
\end{eqnarray}
that $\psi$ is small inclination angle and we set it $\psi=0$. Fig. \ref{fig:ACg} illustrates the variation of redshift--factor in terms of $r/M$ considering $\rho_0 M^2=1$ and $\theta=\pi/8$.
As shown, while at large radii the redshift--factor is independent of King dark matter parameters and approaches $1$, at smaller radii, the influence of these parameters on redshift--factor is such that, for fixed $r/M$ and $\rho_0 M^2$, increasing the size of $R/M$ results in a reduction of the redshift--factor.

\begin{figure}[ht!]
	\centering
  \includegraphics[width=6.4cm]{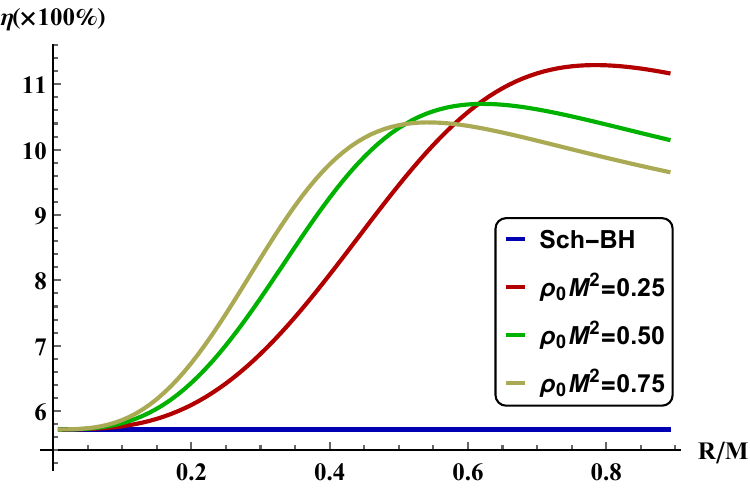} \hspace{-0.2cm}
    \caption{Behavior of the radiative efficiency as $R/M$ varies. The magnitude of this quantity for the Schwarzschild black hole is $\eta\approx 5.72\%$, and the presence of King dark matter increases this value.}\label{fig:AC5}
\end{figure}

Ultimately, a key quantity relevant to the radiation emitted by the accretion disk is the efficiency of the source in transforming the incoming mass into emitted radiation, referred to as $\eta$ and expressed as \cite{kurmanov2022accretion}
\begin{eqnarray}
\eta=\left(1-E(r_{\rm ISCO})\right)\dot{m}.
\end{eqnarray}
Figure \ref{fig:AC5} represents the evolution of this quantity as a function of $R/M$.
It is noteworthy that as $R$ or $\rho_0$ approaches zero, the black hole, in the presence of King dark matter, transforms into a Schwarzschild black hole, with an initial value of $\eta\approx 5.72\%$, reflecting the efficiency of the Schwarzschild black hole. As $R/M$ increases, for small values of $R/M$ the efficiency improves with increasing $\rho_0 M^2$. However, as $R/M$ continues to grow, the magnitude of this quantity becomes larger for smaller values of $\rho_0 M^2$.

\section{Topological Characteristics of Photon Sphere}\label{Sec6}
\begin{figure}[ht!]
	\centering
  \includegraphics[width=6.4cm]{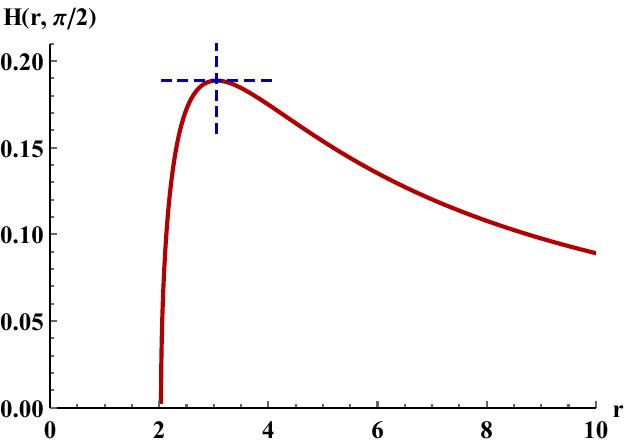} \hspace{-0.2cm}
    \caption{Curve $H(r,\pi/2)$ for $M=1$, $\rho_0=0.5$ and $R=0.1$ at $\theta=\pi/2$. One extreme is at $r_c=3.05240$.}\label{fig:TopoH}
\end{figure}

In Section \ref{Sec3}, we explored the photon sphere of a black hole in the presence of King dark matter and observed that this black hole possesses a photon sphere. However, we have not analyzed the stability or instability of this photon sphere, something that has received considerable attention \cite{qiao2025existence,qiao2022curvatures,koga2019stability}, focusing only on the calculation of the shadow. A stable photon sphere is defined as a region where small perturbations in the light's  path cause neither escape nor capture: the light remains in its orbit.
In contrast, instability means that light is easily disturbed from its circular orbit, ultimately contributing to the formation of the black hole's shadow \cite{qiao2022curvatures,koga2019stability,shoom2017metamorphoses}.
In fact, every black hole has at least one unstable photon sphere that causes the creation of its shadow \cite{cvetivc2016photon,cunha2017fundamental}. Since the distinction between stable and unstable photon spheres is crucial for interpreting observational data on black holes, in this section we investigate the stability or instability of the black hole photon sphere using a topological approach.
For this purpose, a potential is defined as \cite{wei2020topological,sadeghi2024role,sadeghi2024thermodynamic,shahzad2025topological}
\begin{eqnarray}\label{HPot}
H(r,\theta)=\frac{1}{\sin\theta}\sqrt{\frac{f(r)}{h(r)}},
\end{eqnarray} 
showing in Figure \ref{fig:TopoH} its behavior as a function of $r$ (considering $M=1$, $\rho_0=0.5$ and $R=0.1$), where it can be seen that this curve has a maximum located at $r_c=3.05240$.

\begin{figure}[ht!]
	\centering
  \includegraphics[width=6.4cm]{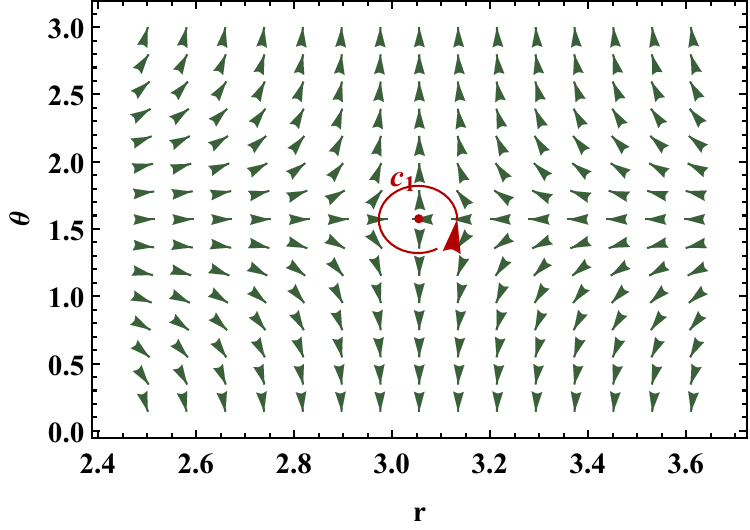} \hspace{0.1cm}
  \includegraphics[width=4.5cm]{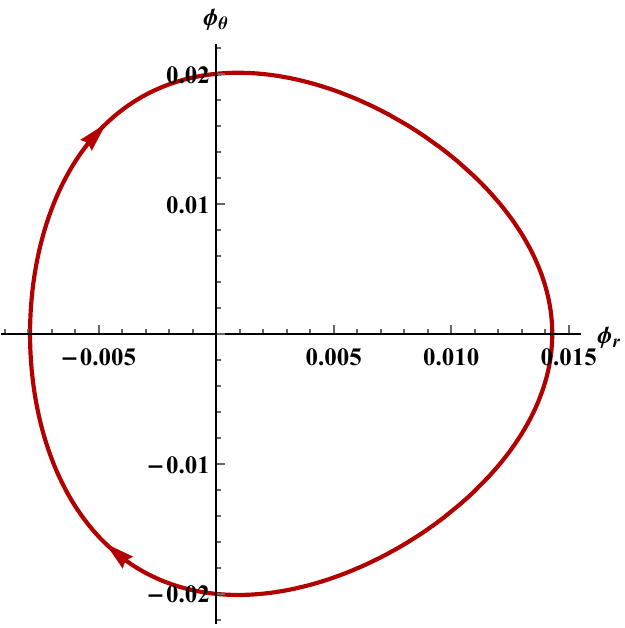} \hspace{0.1cm}
  \includegraphics[width=6.5cm]{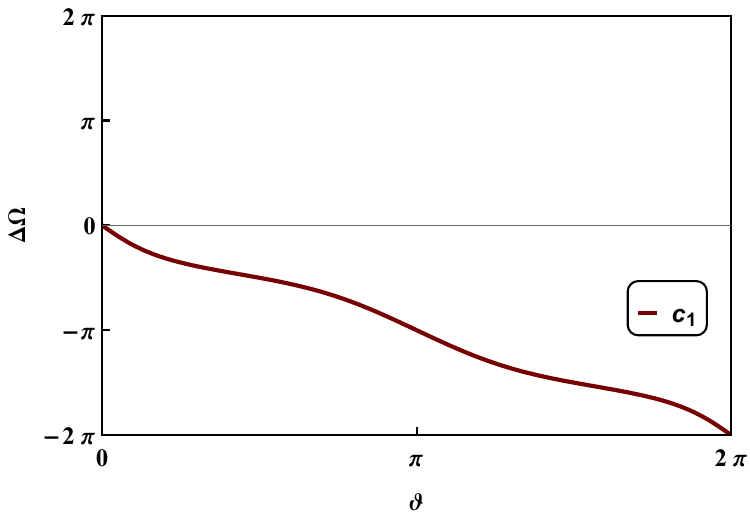} \hspace{-0.1cm}\\
    \caption{
Left panel: Vector space of the potential $H$, with zero point at $r_{c_1}=3.05240$, around which the closed contour $c_1$ is drawn around this zero point. Middle panel: The variation of $\phi_\theta$ as a function of $\phi_r$ around the zero point $r_{c_1}$ assigns a topological charge of $-1$.
Right panel: Variation of $\tilde{\Omega}$ as a function of the parameter $\vartheta$ when setting $a=b=0.3$ for the zero point $r_{c_1}$. This curve indicates that $\omega_1=-1$. In all graphs, the following values have been taken for the parameters: $M=1$, $\rho_0=0.5$ and $R=0.1$.}\label{fig:TopoPS}
\end{figure}

The potential $H$ can be visualized on a plane using coordinates, which are essentially the components of its gradient vector \cite{gashti2025thermodynamic,s2024effective,sekhmani2024thermodynamic}
\begin{align}
\phi_{r}^H =\sqrt{f(r)} \partial_r H(r,\theta),\qquad
\phi_{\theta}^H =\frac{1}{\sqrt{h(r)}}\partial_{\theta} H(r,\theta),
\end{align}
normalized via \cite{wei2020topological,dong2025some,pantig2025multimodal}
\begin{eqnarray}\label{norm}
n^H_r=\frac{\phi^H_{r}}{||\phi||},\qquad
n^H_{\theta}=\frac{\phi^H_{\theta}}{||\phi||},\qquad\text{where}\qquad
||\phi||=\sqrt{(\phi^{H}_r)^2+(\phi^{H}_{\theta})^2}.
\end{eqnarray}
It is evident that at points with coordinates $\theta=\pi/2$, the angular component becomes zero, and the radial component $\partial_r H(r,\theta)\big|_{r=r_c}=0$ also equals zero for these vectors, where the value $r_c$ corresponds to the solution to the equation \eqref{photon} and is essentially where the black hole's photon sphere resides.
These points can be considered `zero points' in the vector space, where vectors converge or diverge, and can be considered topological defects in the vector space, which are assigned a topological number that characterizes the rotation of the vectors around each point.
For zero points, this topological number takes values of  $-1$ or $+1$ depending on the rotations, and is zero at other points. To calculate it, one can examine the closed curve $\phi_r-\phi_\theta$. If the direction of this curve is clockwise for a zero point, the topological number is $-1$, and if it is counterclockwise, it is $+1$ \cite{rizwan2025universal,zhu2025universal,dong2025thermodynamic}.
Another approach is to draw a closed contour around any arbitrary point to analyze the rotation of vectors, and by calculating \cite{wei2020topological,hazarika2024thermodynamic,yerra2022topology}
\begin{eqnarray}\label{winding}
\tilde{\Omega}=w_i=\frac{1}{2\pi}\oint_{c_i} d(\arctan\frac{n_\theta}{n_r}),
\end{eqnarray}
determine the topological number of each point in the vector space. To do this, we employ a parameter change in the form  \cite{wei2020topological,yerra2022topology2}
\begin{eqnarray}
r=a \cos\vartheta +r_i,\qquad
\theta=b \sin\vartheta+\frac{\pi}{2},
\end{eqnarray}
where $r_i$ is the radial coordinate of the point under consideration. 
In Figure \ref{fig:TopoPS} the vector space for the potential $H$ is depicted for the choice  $M=1$, $\rho_0=0.5$, and $R=0.1$.
It is evident that this black hole has only one photon sphere located at $r_{c_1}=3.05240$. The topological charge of this point, considering the clockwise rotation of the $\phi_r-\phi_\theta$ curve and the variations of $\tilde{\Omega}$ with respect to $\vartheta$, is $\omega_1=-1$. 
In the topological analysis of the photon sphere, a charge of $-1$ means its instability \cite{wei2020topological}, which leads to the formation of the black hole shadow, a result consistent with our findings in Section~\ref{Sec3}.

\section{Topological Characteristics of Thermodynamic Potentials}\label{Sec7}

\begin{figure}[ht!]
	\centering
  \includegraphics[width=6.4cm]{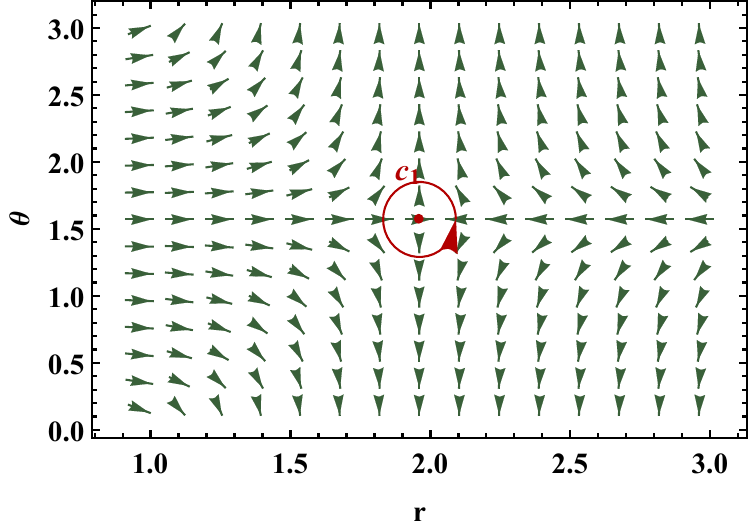} \hspace{0.1cm}
  \includegraphics[width=4.5cm]{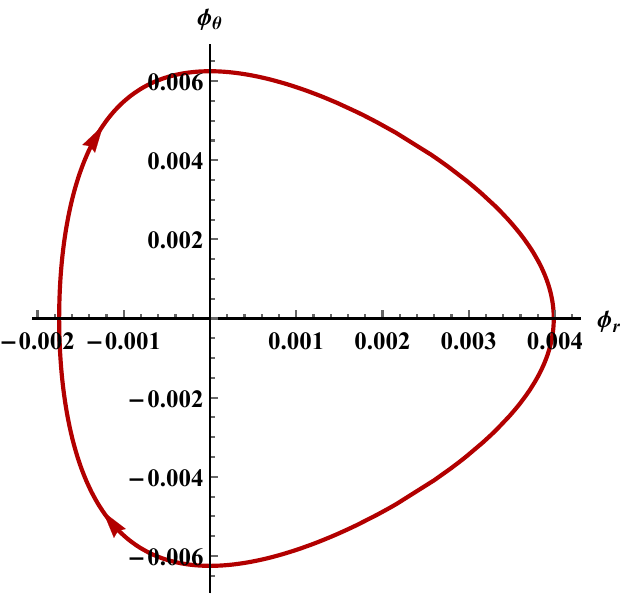} \hspace{0.1cm}
  \includegraphics[width=6.5cm]{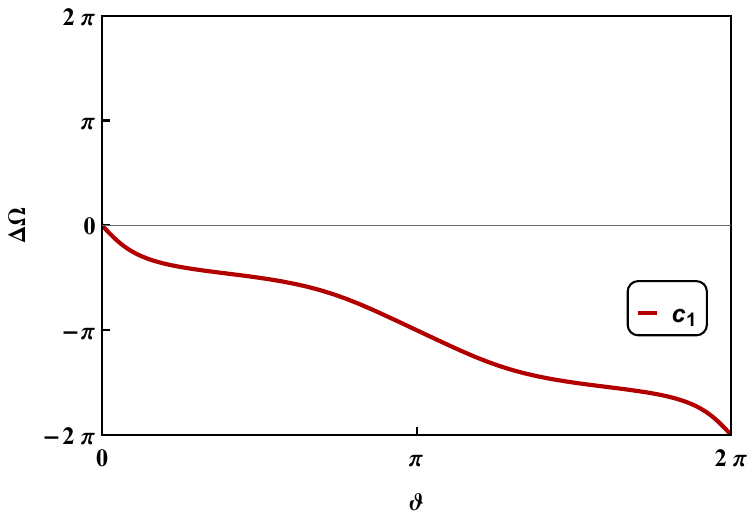} \hspace{-0.1cm}\\
    \caption{Left panel: Vector space of the potential $\Phi$. A zero point is located at $r_{c_1}=1.96087$, enclosed by the closed curve $c_1$. Middle panel: Curve $\phi_r-\phi_\theta$ for the zero point $c_1$. The clockwise direction indicates that $\omega_1=-1$. Right panel: The behavior of $\tilde{\Omega}$ as $\vartheta$ varies for $a=b=0.3$ assigns a topological charge of $-1$. In all cases $\rho_0=1$ and $R=0.35$.}\label{fig:TopoTemp}
\end{figure}

We studied the Hawking temperature of the black hole in Section \ref{Sec2} and observed that there is a phase transition for some selected parameters in its curve at $\partial_{r_h}T_{\rm H}\big|_{r_h=r_c}=0$, but we did not analyze the nature of this phase transition. In this section, building on the findings of the previous section, we will use a topological approach to examine this phase transition and identify the type of critical point involved. 
In addition, we will delve deeper into the  analysis of the phase transitions present in the generalized free energy of the black hole outside the horizon shell. In doing so, we will also determine the topological class to which the black hole in Eq. \eqref{ds2} belongs.
Therefore, we define a potential dependent on the Hawking temperature of the black hole in the form of \cite{wei2022topology,barzi2024renyi,chen2024thermal,alipour2023topological}
\begin{eqnarray}
\Phi =\frac{1}{\sin \theta} T_{\rm H},
\end{eqnarray}
which is represented in vector space using the vectors \cite{yerra2023topology,mehmood2023thermodynamic,zhang2023topology}
\begin{equation}
\phi^{\Phi}_r=\partial_{r_h}\Phi,\qquad
\phi^{\Phi}_\theta=\partial_\theta\Phi,
\end{equation}
which are normalized similarly to the equation \eqref{norm}. It is evident that the zero points of this vector space are located at $\theta=\pi/2$ and $\partial_{r_h}T_{\rm H}\big|_{r_h=r_c}=0$. As mentioned above in section \ref{Sec2}, when analyzing black holes thermodynamically, we are interested in states with remnants. In these states, where the parameters are selected according to Figure \ref{fig:rRem}, the Hawking temperature has only one extreme.
The vector space of the potential $\Phi$ is illustrated for the selection of $\rho_0=1$ and $R=0.35$ in Fig. \ref{fig:TopoTemp}.
As expected, there is only one zero point at $r_{c_1}=1.96087$, whose topological number, considering the rotation of the  $\phi_r-\phi_\theta$ curve and the variations of $\tilde{\Omega}$ with respect to $\vartheta$, is $-1$.
The topological number of $\omega_1=-1$ indicates that the critical point $r_c$ is of the conventional type \cite{wei2022topology}. Thus, we can conclude that, by choosing the parameters $R$ and $\rho_0$ according to the color region in Fig. \ref{fig:rRem}, a conventional critical point exists at the Hawking temperature of the black hole.

\begin{figure}[ht!]
	\centering
  \includegraphics[width=6.1cm]{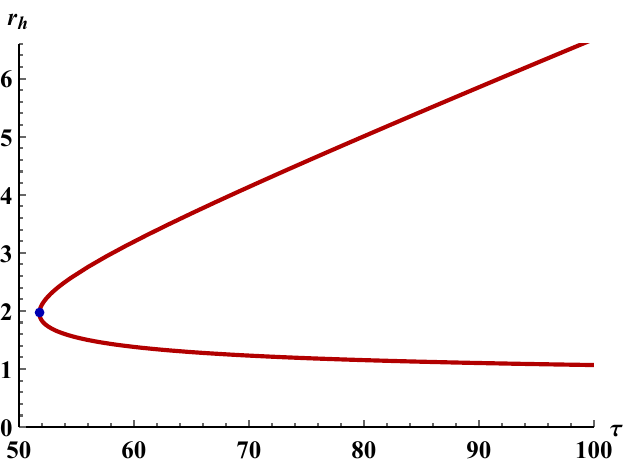}
    \caption{Curve $r_h-\tau$ for $\rho_0=1$ and $R=0.35$. This curve changes direction at $\tau_c=51.79746$ and has two branches for $\tau>\tau_c$.}\label{fig:TopoTau}
\end{figure}

\begin{figure}[ht!]
	\centering
  \includegraphics[width=6.4cm]{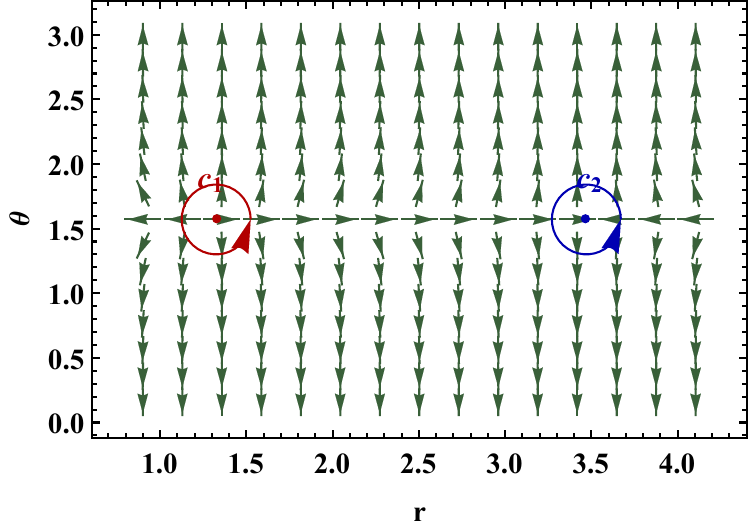} \hspace{0.1cm}
  \includegraphics[width=4.5cm]{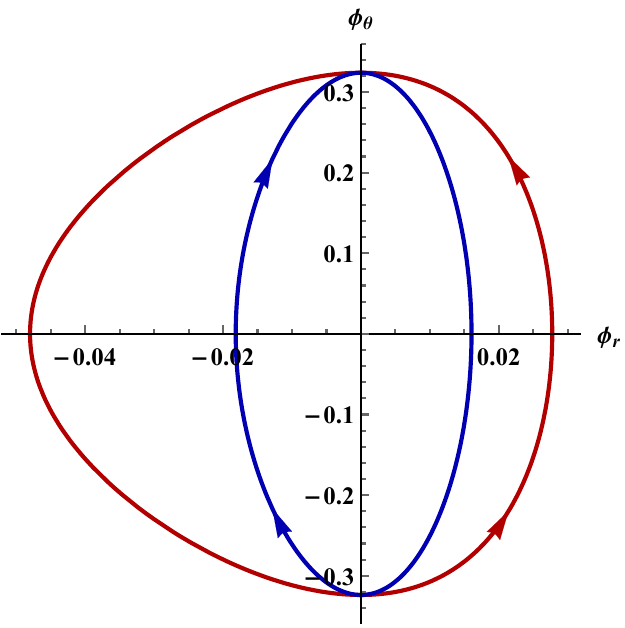} \hspace{0.1cm}
  \includegraphics[width=6.5cm]{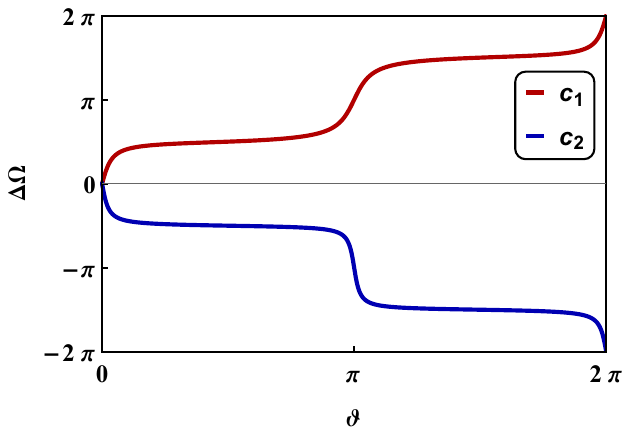} \hspace{-0.1cm}\\
    \caption{Assuming $\rho_0=1$, $R=0.35$, and $\tau=20\pi$. Left panel: Vector space of potential $\mathcal{F}$ that contains two zero points $r_{c_1}=1.32501$ and $r_{c_2}=3.47124$ which are enclosed by the closed contours $c_1$ and $c_2$, respectively. Middle panel: The direction of $\phi_r-\phi_\theta$ curve for $r_{c_1}$ and $r_{c_2}$ indicates a topological charge of $+1$ and $-1$, respectively. Right panel: $\tilde{\Omega}$ curve in terms of $\vartheta$ for closed contours $c_1$ and $c_2$ refers to a topological charge of $\omega_1=+1$ and $\omega_2=-1$, respectively.}\label{fig:TopoHelm}
\end{figure}

The generalized free energy of the black hole can also be studied in a similar way. This energy, outside the horizon, is given by \cite{wei2022black,wu2023topological,gashti2024topology,di2024topological,liu2024thermodynamic,hosseinifar2025quasinormal}
\begin{equation}
\mathcal{F}=M(r_h)-\frac{S}{\tau}
\end{equation}
where $\tau$ is the inverse temperature outside the horizon and $S$ represents the entropy of the black hole, expressed by the equations \eqref{Mh} and \eqref{TH} as \cite{bekenstein1973black,de2012black}
\begin{eqnarray}
S=\int\frac{d M(r_h)}{T_H} dr_h=\pi r_h^2
\end{eqnarray}
which corresponds to the Schwarzschild black hole entropy.
The vector space of this potential can be represented using the vectors \cite{wu2023topological2,wu2025novel,azreg2025thermodynamic,rizwan2023topological}
\begin{equation}
\phi_r^{\mathcal{F}}=\partial_{r_h}\mathcal{F},\qquad
\phi_\theta^{\mathcal{F}}=-\cot \theta \csc \theta.
\end{equation}
normalized according to Eq. \eqref{norm}.
The zero points of this potential are located at $\theta=\pi/2$ and $\partial_{r_h}\mathcal{F}\big|_{r=r_c}=0$. Furthermore, using $\partial_{r_h}\mathcal{F}=0$, a relation for $\tau$ in terms of $r_h$ can be derived, expressed as
\begin{eqnarray}
\tau=\frac{4 \pi  r_h \left(r_h^2+R^2\right)^{3/2}}{\left(r_h^2+R^2\right)^{3/2}-8 \pi  \rho_0 r_h^2 R^3}.
\end{eqnarray}
Fig. \ref{fig:TopoTau} illustrates the variations of $r_h$ with respect to $\tau$ for $\rho_0=1$ and $R=0.35$, 
showing that at $\tau_c=51.79746$, this curve changes direction and has two branches for $\tau>\tau_c$.
In this way, by selecting an appropriate $\tau$, the vector space of this potential can be represented.
Figure \ref{fig:TopoHelm} shows the vector space of the potential $\mathcal{F}$ for $\rho_0=1$, $R=0.35$, and $\tau=20\pi$.
It is evident that this vector space has two zero points at $r_{c_1}=1.32501$ and $r_{c_2}=3.47124$, whose topological numbers, based on the direction of the  $\phi_r-\phi_\theta$ curve and the variations of $\tilde{\Omega}$ with respect to $\vartheta$, are $\omega_1=+1$ and $\omega_2=-1$ for the closed curves $c_1$ and $c_2$, respectively. Therefore, the total topological charge for this potential is zero, leading us to conclude that this black hole lies within the topological class similar to Reissner-Nordstr{\"o}m black hole \cite{wei2022black}.


\section{Conclusions}\label{Sec8}

In this work, we have carried out a thorough investigation of the physical and thermodynamic properties of a Schwarzschild black hole surrounded by King-type dark matter. Our analysis shows that the inclusion of the King dark matter parameters, $R$ and $\rho_{0}$, introduces significant deviations from the standard Schwarzschild case.
In particular, when any of these parameters tends to zero, the system gradually recovers the Schwarzschild solution, confirming the consistency of the model. Conversely, increasing any of the parameters causes the radius of the event horizon to shift outward, highlighting the role of dark matter in effectively increasing the gravitational influence of the central object.

The thermodynamic behavior of the system was analyzed by studying the Hawking temperature. It was observed that, for specific parameter ranges, the temperature exhibits a single extreme, implying the existence of a remnant radius below which the black hole cannot continue evaporating. 
This feature points to the possibility of stable remnants induced by dark matter, which could be of interest for addressing questions related to black hole evaporation and the endpoint of Hawking radiation. A viable parameter space consistent with such remnants has been identified.

We further explored the propagation of massive and massless particles in this background. For photons, we determined the  radius of the black hole shadow and established that its size increases monotonically with either $R$ or $\rho_{0}$, while returning to the Schwarzschild value when both vanish. By comparing these results with EHT observations, we derived upper bounds for the dark matter parameters that ensure consistency with current astrophysical constraints. 
For timelike geodesics, we examined the dynamics of massive particles by calculating the ISCO, as well as the associated angular velocity, angular momentum, and specific energy. The ISCO radius was found to increase with dark matter parameters, while the angular momentum and angular velocity increased  compared to the Schwarzschild case, accompanied by a corresponding reduction in particle energy. These deviations reflect the impact of the surrounding dark matter on accretion processes around black holes.

To better understand the astrophysical signatures, we calculated the energy flux of the accretion disk and subsequently derived the effective radiation temperature, differential luminosity, and spectral luminosity. The qualitative behavior of the radiation temperature and differential luminosity was found to closely follow the energy flux profile, while the presence of King's dark matter influenced both the spectral luminosity and  radiative efficiency. These modifications could serve as potential observational signatures of dark matter halos in black hole environments.

From the perspective of photon dynamics, we also studied the stability of the photon sphere in the presence of King's dark matter. Our analysis revealed the existence of a unique unstable photon sphere, which directly contributes to the shadow formation and remains consistent with the expected causal structure of spacetime.

Finally, we turned to the thermodynamic aspects of the black hole using a topological approach. We identified a conventional critical point in the Hawking temperature that coincides with the parameter regimes giving rise to a remnant radius. The study of the off-shell generalized free energy further revealed the occurrence of two distinct phase transitions. A topological classification of the solution indicates that the Schwarzschild black hole, surrounded by King dark matter, belongs to the same universality class as the 
Reissner-Nordstr\"om black hole, 
establishing a deeper 
connection between the two systems.

Our results underscore the significant role of the King dark matter model in altering the geometry, dynamics, and thermodynamics of black holes. Modifications in the horizon structure, particle trajectories, accretion disk properties, and thermodynamic phase behavior provide a robust framework for investigating the effects of dark matter  in strong gravity regimes. 
These findings could have observational consequences, particularly in the interpretation of black hole shadows, accretion disk spectra, and potential remnant states. Future research could extend this analysis to rotating spacetimes, explore alternative dark matter profiles, or investigate possible signatures in gravitational wave observations, thus offering a more complete understanding of the interplay between black holes and dark matter in astrophysical contexts.


\section*{Acknowledgements}
The research of L.M.N., S.Z. and H.H. was supported by the Q-CAYLE project, funded by the European Union-Next Generation UE/MICIU/Plan de Recuperacion, Transformacion y Resiliencia/Junta de Castilla y Leon (PRTRC17.11), and also by project PID2023-148409NB-I00, funded by MICIU/AEI/10.13039/501100011033. Financial support of the Department of Education of the Junta de Castilla y Leon and FEDER Funds is also gratefully acknowledged (Reference: CLU-2023-1-05).
D.J.G. acknowledges the contribution of the COST Action CA21136  -- ``Addressing observational tensions in cosmology with systematics and fundamental physics (CosmoVerse)".
Additionally, H. H. is grateful to Excellence project FoS UHK 2203/2025-2026 for the financial support.



\end{document}